\documentclass[conference,a4paper]{SISA2021}
\usepackage{multirow}
\usepackage[pdftex]{graphicx}
\usepackage{amsmath}
\usepackage{amsxtra}
\usepackage{threeparttable}
\usepackage{cite}
\usepackage{hyperref}
\usepackage{enumitem}
\usepackage{subfigure}
\usepackage{xcolor}

\begin{document}

\title{Protecting Semantic Segmentation Models by Using Block-wise Image Encryption with Secret Key from Unauthorized Access}

\author{%
\IEEEauthorblockN{%
Hiroki Ito,
MaungMaung AprilPyone, and
Hitoshi Kiya}
\IEEEauthorblockA{%
Tokyo Metropolitan University, Japan}}
%
%
%

\maketitle
\thispagestyle{empty}

\begin{abstract}
  Since production-level trained deep neural networks (DNNs) are of a great business value, protecting such DNN models against copyright infringement and unauthorized access is in a rising demand. However, conventional model protection methods focused only the image classification task, and these protection methods were never applied to semantic segmentation although it has an increasing number of applications. In this paper, we propose to protect semantic segmentation models from unauthorized access by utilizing block-wise transformation with a secret key for the first time. Protected models are trained by using transformed images. Experiment results show that the proposed protection method allows rightful users with the correct key to access the model to full capacity and deteriorate the performance for unauthorized users. However, protected models slightly drop the segmentation performance compared to non-protected models.
\end{abstract}

\section{Introduction}
\label{sec:intro}
In recent years, deep neural networks (DNNs) have been actively studied in many areas such as computer vision (CV) and natural language processing, and their performance has been greatly improved \cite{dl}.
Especially in CV domain, the performance of DNNs is approaching the practical level not only for a simple classification task but also for advanced tasks such as object detection and segmentation.
Semantic segmentation aims to recognize what is in the image at pixel level and has an increasing number of applications such as handwritten text recognition, automated driving, and medical image analysis \cite{application1, application2, application3}.

However, training a model with high performance is generally very expensive because it requires huge data, powerful computing resources, and human expertise. 
For example, ImageNet, which has more than 14 million images in over 20,000 categories, was created using much effort with crowdsourcing to ensure that images were correctly labeled \cite{imagenet}.
In addition, training on such a large dataset consumes days and weeks even on GPU-accelerated machines.
Therefore, a model trained at the production level has a great value and should be treated as a valuable intellectual property (IP) considering its training costs.

In the research to protect trained models, there are two types of concepts, namely ownership verification and access control.
Ownership verification aims to reveal the ownership of a model.
The study in \cite{embed1} proposed to embed a watermark in a model without accuracy degradation.
In \cite{embed2}, the authors leave a backdoor by embedding watermarks in parts of the training data and assigned them into new labels.
However, unauthorized users, such as the attacker who stole the model, can use the model adequately same as authorized users without arousing any suspicion.
In addition, the stolen model can be exposed to a variety of attacks, for instance, model inversion attacks \cite{inv-attack} and adversarial attacks \cite{adv-attack}.
Access control aims that authorized users can fully utilize the benefits of the model, while unauthorized users are not able to use the model in full capacity.
In this paper, we focus on access control and target to prevent unauthorized users from using the model correctly even if the model is stolen.

In recent studies on access control, authorized users perform pre-processing on the input image, such as adding a noise or block-based transformation \cite{access1, access2, access3}.
The work in \cite{access1} is inspired by adversarial examples \cite{adv-attack}, and perturbations need to be added to the input images using a transformation module to obtain correct predictions.
The study in \cite{access2, access3} introduced a secret key for protecting a model, and the method was inspired by adversarial examples and image encryption \cite{enc3, perceptual1}.
The secret key-based protection method \cite{access3} uses a key-based transformation that was originally used by an adversarial defense in \cite{adv-def}, which was in turn inspired by perceptual image encryption methods \cite{perceptual1, perceptual2, perceptual4, perceptual6, perceptual7}.
This model protection method utilizes a secret key in such a way that a stolen model cannot be used to its full capacity without a correct secret key.
However, these works were evaluated only on the classification task, and it is not known how well they perform on other advanced tasks.

Therefore, in this paper, we propose to protect semantic segmentation models from unauthorized access by utilizing block-wise transformation with a secret key for the first time.
In our scenario, training images are block-based transformed, and segmentation models are trained using the transformed images and the corresponding ground truth.
The transformation methods are lightweight processing and do not require any modifications to the model.
Experiment results show that the protection method allows rightful users with the correct key to access the model to full capacity and deteriorate the performance for unauthorized users. However, protected models slightly drop the segmentation performance compared to non-protected models.

\section{Protecting Semantic Segmentation Models}
\label{sec:proposed}
\subsection{Overview}
\begin{figure*}[t!]
	\centering
 	\centerline{\includegraphics[width=16cm] {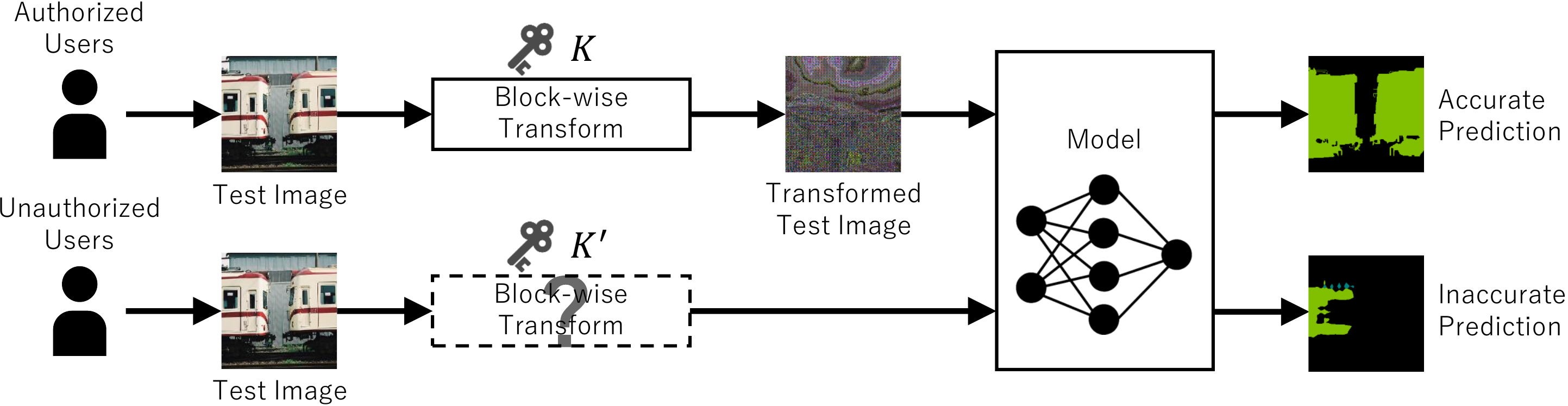}}
  	\caption{Framework for protecting semantic segmentation models}
	\label{fig:framework}
\end{figure*}

Figure \ref{fig:framework} illustrates the framework for protecting semantic segmentation models from unauthorized access.
In this framework, authorized users transform test images by using a block-wise transformation with correct secret key $K$ and input the transformed images into a model.
The model predicts the transformed images correctly with high accuracy.
In contrast, unauthorized users are not able to transform test images correctly because they do not know a necessary transformation or correct secret key $K$ (i.e., using an incorrect key $K'$).
Therefore, the model returns inaccurate predictions to the unauthorized users.

\subsection{Semantic Segmentation}
\begin{figure}[t]
	\centering
 	\centerline{\includegraphics[width=8.5cm] {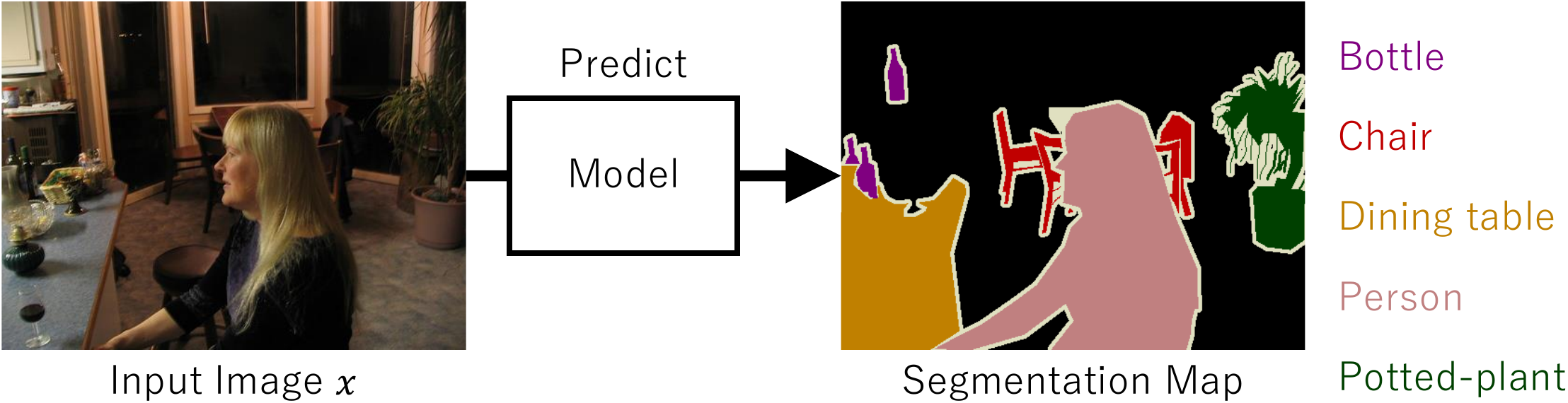}}
  	\caption{Example of semantic segmentation}
	\label{fig:semantic}
\end{figure}

The goal of semantic segmentation is to understand what is in an image at the pixel level.
Figure \ref{fig:semantic} shows an example of semantic segmentation.
The segmentation model takes an input image $x$ with the shape $c \times h \times w$ and outputs a segmentation map (SM) with the shape $1 \times h \times w$.
Here, $c$, $h$, and $w$ are the number of channels, height, and width of $x$, respectively, and each pixel in the SM represents a class label.

One of the common metrics to evaluate segmentation performance is mean intersection-over-union (mean IoU) \cite{fcn, giou}, which first computes the IoU for each class and then computes the average over classes.
The IoU metric is the area of overlap between a predicted SM and a corresponding ground truth (GT) divided by the area of the union.
Mean IoU is defined as
\begin{equation}
    \mathrm{mean IoU}(SM, GT) = \frac{1}{class} \sum_{i=1}^{class} \frac{|SM_{i} \cap GT_{i}|}{|SM_{i} \cup GT_{i}|},
\end{equation}
where $class$ is the number of classes.
In addition, the metric ranges from zero to one, where a value of one means that a SM and a GT overlap perfectly, and a value of zero indicates no overlap.

\subsection{Block-wise transformation Procedure}
\begin{figure*}[t!]
	\centering
 	\centerline{\includegraphics[width=16cm] {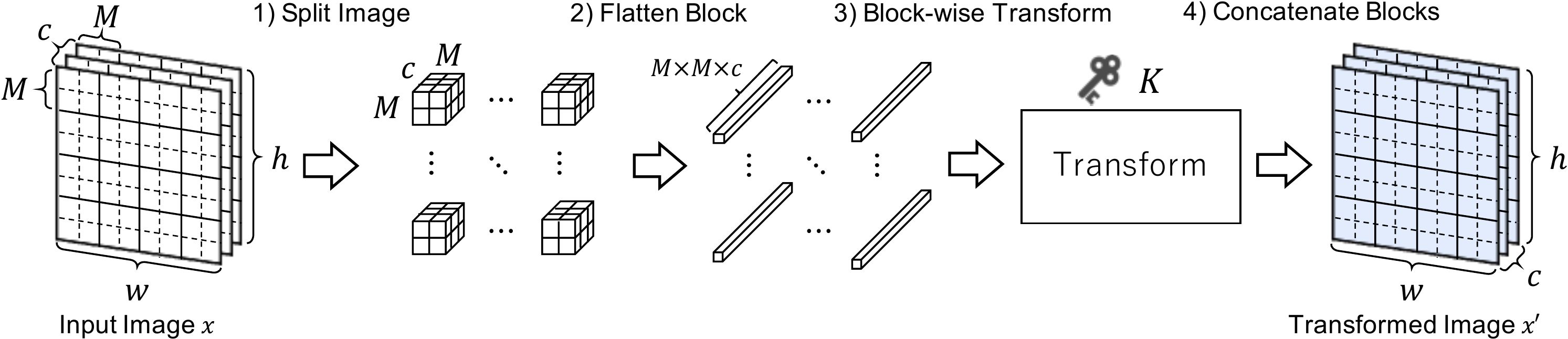}}
  	\caption{Block-wise transformation process}
	\label{fig:transform}
\end{figure*}

In this paper, we used block-wise transformation with a secret key for access control of semantic segmentation models.
The transformation was proposed in \cite{adv-def} for adversarial defense.
We perform the block-wise transformation in the following steps (see Fig.\ \ref{fig:transform}).
\begin{enumerate}
    \item Split Image: Segment an input image $x$ with the shape $c \times h \times w$ into blocks. Each block has a shape of $c \times M \times M$, where $M$ is the block size.
    \item Flatten Block: Flatten each block into a vector of length $c \times M \times M$.
    \item Block-wise Transform: Each flattened block is transformed using a block-wise transformation with provided secret key $K$. All flattened blocks are converted with the same key.
    \item Concatenate Blocks: The transformed blocks are integrated in the reverse order of the block splitting to obtain the transformed image $x'$.
\end{enumerate}

We used three transformation methods for the block-wise transformation. The details of these methods are described in the next section.

\subsection{Transformation Methods}
\begin{figure*}[t!]
    \centering
    \begin{tabular}{c|ccccc}
    \multirow{2}{*}{Transformation} & \multicolumn{5}{c}{Block size $M$} \\ 
     & 2 & 4 & 8 & 16 & 256 \\ \hline \\[-5pt]
    SHF & 
    \begin{minipage}{2.5cm}
      \centering
      \includegraphics[width=2.5cm]{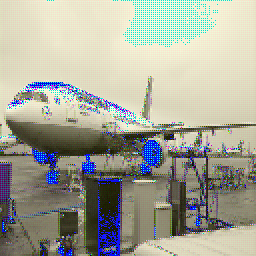}
    \end{minipage} &
    \begin{minipage}{2.5cm}
      \centering
      \includegraphics[width=2.5cm]{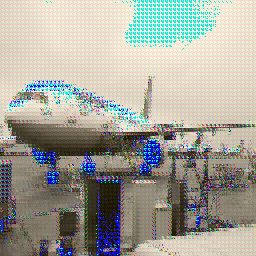}
    \end{minipage} &
    \begin{minipage}{2.5cm}
      \centering
      \includegraphics[width=2.5cm]{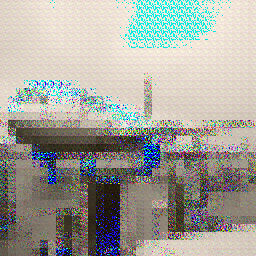}
    \end{minipage} &
    \begin{minipage}{2.5cm}
      \centering
      \includegraphics[width=2.5cm]{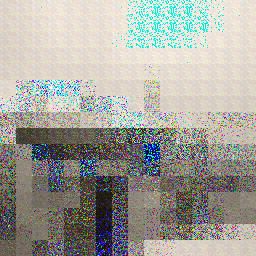}
    \end{minipage} &
    \begin{minipage}{2.5cm}
      \centering
      \includegraphics[width=2.5cm]{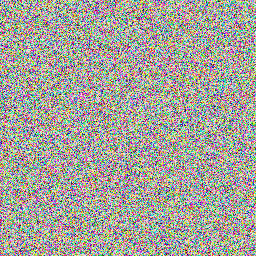}
    \end{minipage} \\ \\[-5pt]
    NP & 
    \begin{minipage}{2.5cm}
      \centering
      \includegraphics[width=2.5cm]{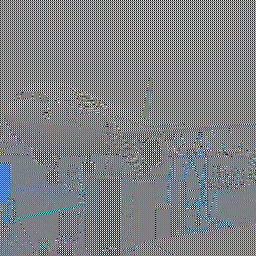}
    \end{minipage} &
    \begin{minipage}{2.5cm}
      \centering
      \includegraphics[width=2.5cm]{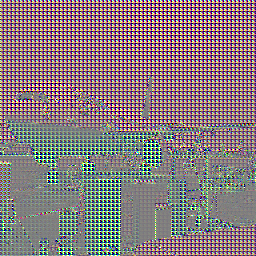}
    \end{minipage} &
    \begin{minipage}{2.5cm}
      \centering
      \includegraphics[width=2.5cm]{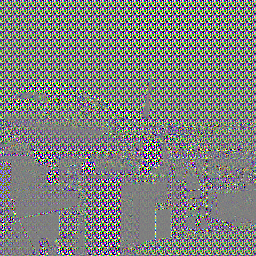}
    \end{minipage} &
    \begin{minipage}{2.5cm}
      \centering
      \includegraphics[width=2.5cm]{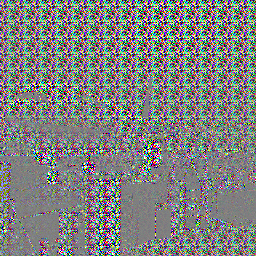}
    \end{minipage} &
    \begin{minipage}{2.5cm}
      \centering
      \includegraphics[width=2.5cm]{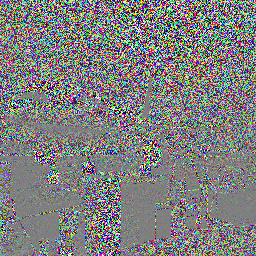}
    \end{minipage} \\ \\[-5pt]
    FFX & 
    \begin{minipage}{2.5cm}
      \centering
      \includegraphics[width=2.5cm]{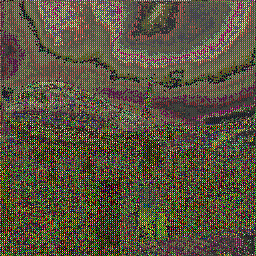}
    \end{minipage} &
    \begin{minipage}{2.5cm}
      \centering
      \includegraphics[width=2.5cm]{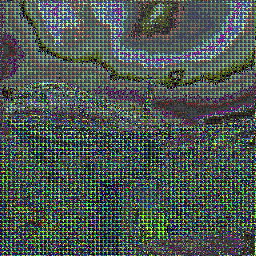}
    \end{minipage} &
    \begin{minipage}{2.5cm}
      \centering
      \includegraphics[width=2.5cm]{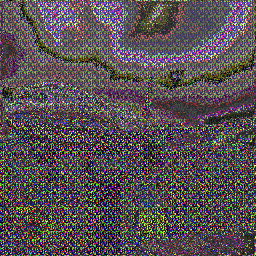}
    \end{minipage} &
    \begin{minipage}{2.5cm}
      \centering
      \includegraphics[width=2.5cm]{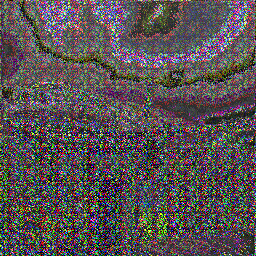}
    \end{minipage} &
    \begin{minipage}{2.5cm}
      \centering
      \includegraphics[width=2.5cm]{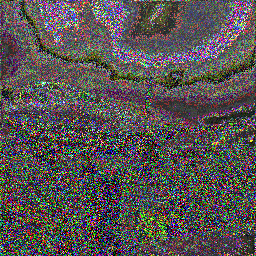}
    \end{minipage} \\
    \end{tabular}
    \caption{Images transformed by block-wise transformation}
    \label{fig:enc-sample}
\end{figure*}

We utilize three transformation methods, namely, pixel shuffling (SHF), negative/positive transformation (NP), and format-preserving Feistel-based encryption (FFX) \cite{ffx}, in step 3 (see Fig.\ \ref{fig:transform}).
Figure \ref{fig:enc-sample} shows a sample of images transformed by using the methods.
\begin{description}[style=unboxed,leftmargin=0cm]
    \item[SHF] shuffles the values in the flattened block on the basis of secret key $K$.
    \item[NP] inverts selected values from the flattened block (i.e., subtract a value from 255) in accordance with $K$.
    In this paper, we apply NP to half of the pixels in a block.
    \item[FFX], with a length of 3 digits, encrypts half of the pixels in the block to cover the whole range from 0 to 255, and the pixels are chosen by using $K$. The pixel values of the input image are in the range from 0 to 255, whereas the encrypted pixel values become in the range from 0 to 999. Then, all pixel values are divided by the maximum pixel value after the encryption to be standardized. In addition, FFX requires a password for format-preserving Feistel-based encryption, and we used a fixed password in this paper.
\end{description}

\section{Experiments and Results}
\label{sec:experiments}
In this section, we conduct experiments to validate the effectiveness of the proposed framework (see Fig.\ \ref{fig:framework}) for protecting semantic segmentation models.

\subsection{Experimental Setup}
We used a fully convolutional network (FCN) \cite{fcn} with a ResNet-50 \cite{resnet} backbone as a semantic segmentation model.
FCN uses a backbone (e.g., ResNet-50) to extract features from an input image and generates a segmentation map by upsampling the output of the backbone to the same size (i.e., height and width) as the input image using bilinear interpolation.
We trained the network by using the PASCAL visual object classes segmentation dataset in 2012 \cite{pascal}.
The dataset consists of a training set with 1464 images and corresponding ground truths, a validation set with 1449 images, and a test set.
However, we tested the model's performance with the validation set because the test set is only available on the evaluation server and has limitations such as the maximum number of submissions.
The training set was divided into 1318 and 146 samples, and we used them for training and validation, respectively.
Since the block-wise transformation requires an input image size to be determined in advance, we resized the height and width of all images and masks to 256.
In addition, standard data-augmentation methods, i.e., random resized crop and horizontal flip, were performed in the training.

All networks were trained for 100 epochs by using stochastic gradient descent with a weight decay of 0.005 and a momentum of 0.9. 
The learning rate was initially set to 0.1 and scheduled by cosine annealing with warm restarts \cite{lr-schedule}.
Here, the minimum learning rate was 0.0001, and the learning rate was restarted every ten epochs.
The batch size was 256.
We used the cross-entropy loss to calculate a loss.
After the training, we selected the model that provided the lowest loss value under the validation.
In addition, we evaluated the segmentation performance by using mean IoU.

\subsection{Segmentation Performance}
To evaluate the effect of the block-wise transformations on segmentation performance, we trained models by using images transformed by various transformations with different block sizes.
Table \ref{tab:accuracy} shows the experimental result, where ``SHF,'' ``NP,'' and ``FFX'' indicate that the training and test images were transformed by the respective methods, and ``Baseline'' means that all images were not transformed.
Example of input images, the ground truth (GT), and the predictions (i.e., segmentation maps) of the models with $M=2$ are shown in Fig.\ \ref{fig:prediction}.
The numbers below the prediction maps indicate the mean IoU values between the ground truth and the prediction maps.

From Table \ref{tab:accuracy}, the segmentation accuracy of SHF dropped significantly as the block size was increased.
NP and FFX also decreased the accuracy when the block size was increased, but the loss of accuracy was less than that of SHF.
The reason why SHF is sensitive to the block size may be that location information is critical in the segmentation task.
In addition, all protected models slightly dropped the segmentation performance compared to the baseline, even in the most accurate conditions ($M=2$).

\begin{table}[t!]
	\centering
	\caption{Segmentation accuracy (mean IoU) \\ of protected models and baseline model}
	\begin{tabular}{c|ccc|c} \hline \hline
		Block size $M$ & SHF & NP & FFX & Baseline \\ \hline
		2   & 0.6597 & 0.6216 & 0.5271 & \multirow{8}{*}{0.7080}\\
        4   & 0.5941 & 0.5695 & 0.4944 & \\
        8   & 0.4723 & 0.5173 & 0.4651 & \\
        16  & 0.2958 & 0.4329 & 0.3444 & \\
        32  & 0.1575 & 0.3880 & 0.2763 & \\
        64  & 0.0693 & 0.3733 & 0.2628 & \\
        128 & 0.0600 & 0.3218 & 0.2417 & \\
        256 & 0.0380 & 0.3007 & 0.2367 & \\ \hline \hline
	\end{tabular}
	\label{tab:accuracy}
\end{table}

\bgroup
\setlength{\tabcolsep}{5pt}
\begin{figure}[t]
    \centering
    \begin{tabular}{c|ccc}
    Method & Input & Prediction & GT \\ \hline \\[-8pt]
    Baseline & 
    \begin{minipage}{2cm}
      \centering
      \includegraphics[width=2cm]{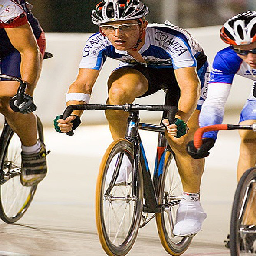}
    \end{minipage} &
    \begin{minipage}{2cm}
      \centering
      \includegraphics[width=2cm]{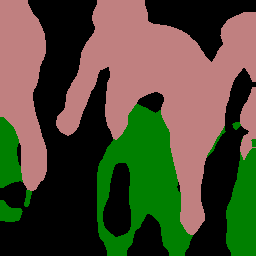}
    \end{minipage} &
    \multirow{8}{2cm}[-2.3cm]{
    \begin{minipage}{2cm}
      \centering
      \includegraphics[width=2cm]{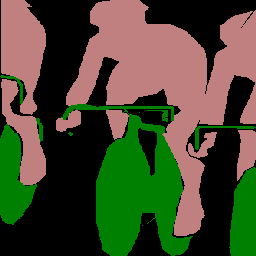}
    \end{minipage}
    } \\
     & & 0.8094 & \\
    SHF & 
    \begin{minipage}{2cm}
      \centering
      \includegraphics[width=2cm]{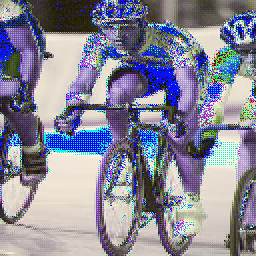}
    \end{minipage} &
    \begin{minipage}{2cm}
      \centering
      \includegraphics[width=2cm]{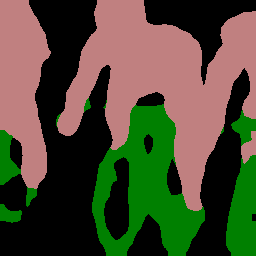}
    \end{minipage} & \\
     & & 0.7745 & \\
    NP & 
    \begin{minipage}{2cm}
      \centering
      \includegraphics[width=2cm]{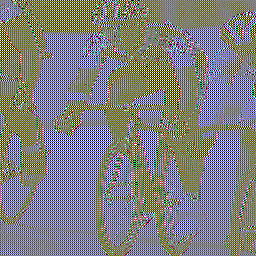}
    \end{minipage} &
    \begin{minipage}{2cm}
      \centering
      \includegraphics[width=2cm]{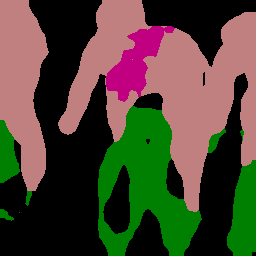}
    \end{minipage} & \\
     & & 0.5506 & \\
    FFX & 
    \begin{minipage}{2cm}
      \centering
      \includegraphics[width=2cm]{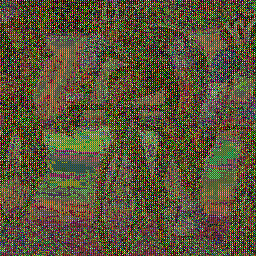}
    \end{minipage} &
    \begin{minipage}{2cm}
      \centering
      \includegraphics[width=2cm]{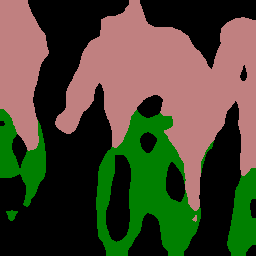}
    \end{minipage} & \\
     & & 0.7142 & \\
    \end{tabular}
    \caption{Example of input test images, predictions from models, and corresponding ground truth. Mean IoU values are given under prediction maps.}
    \label{fig:prediction}
\end{figure}
\egroup

\subsection{Access Control Performance}
\begin{figure}[t]
    \centering
    \subfigure[SHF]{\includegraphics[width=8cm]{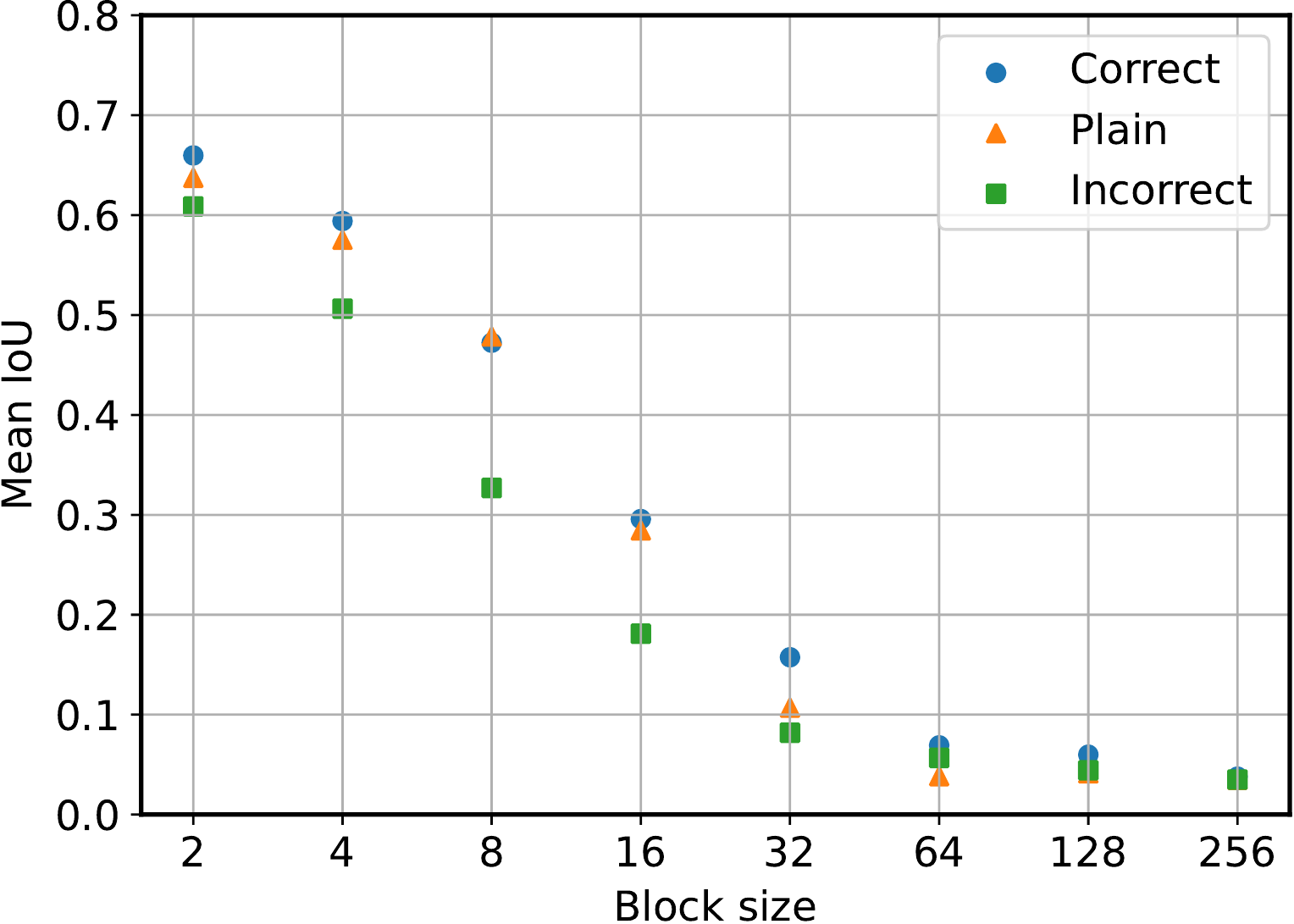}} \\
    \subfigure[NP]{\includegraphics[width=8cm]{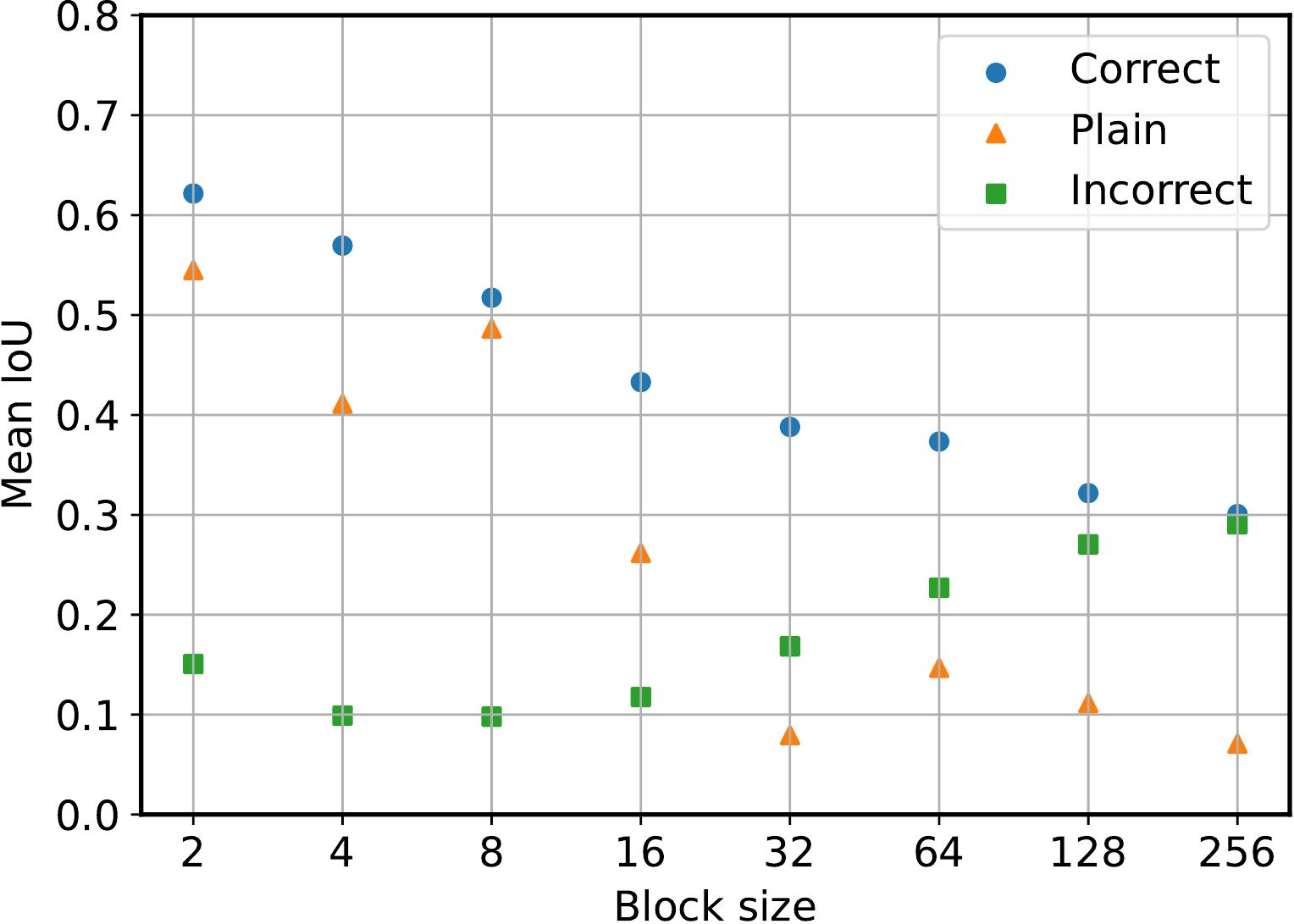}} \\
    \subfigure[FFX]{\includegraphics[width=8cm]{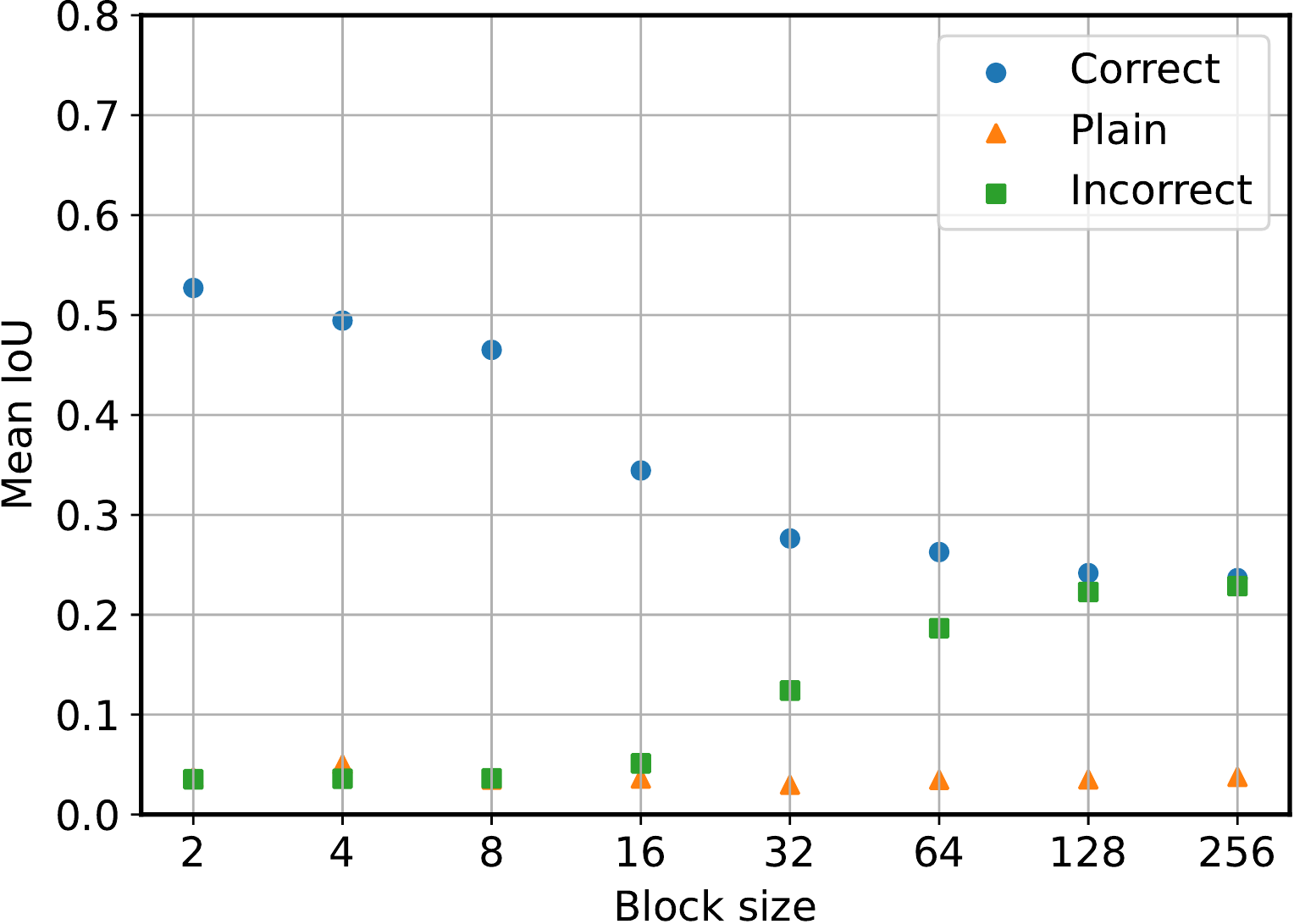}}
    \caption{Access control performance}
    \label{fig:access-control}
\end{figure}

In this section, we tested protected models by using plain images and images transformed with incorrect key $K'$ (i.e., a different key from a correct one used in the model training) to verify access control performance.
The protected models were trained with images transformed by each transformation method (i.e., SHF, NP, and FFX) with correct key $K$.
Figure~\ref{fig:access-control} shows the experimental result, where ``Plain'' and ``Incorrect'' indicate that the models were tested by using plain images and images transformed with $K'$, respectively, and ``Correct'' refers that the models were tested by using images transformed with $K$ (same as Table \ref{tab:accuracy}).
For ``Incorrect,'' the results were averaged over 100 incorrect keys.

As shown in Fig.~\ref{fig:access-control}, among three transformations, FFX had the best access control performance, however, it had a lower segmentation accuracy.
Although SHF and NP had a high segmentation accuracy for small block sizes, the models still worked well for unauthorized access (i.e., using plain images and transformed ones with $K'$), indicating a lower access control performance.
Therefore, there is a trade-off between segmentation accuracy and access control performance.
 

\section{conclusion}
\label{sec:conclusion}
We proposed a method to protect models from unauthorized access by using block-wise transformations, namely pixel shuffling (SHF), negative/positive transformation (NP), and format-preserving Feistel-based encryption (FFX).
In particular, we evaluated the access control performance of the transformations in semantic segmentation, which has not been tested in previous studies.
In experiments for segmentation performance, the highest accuracy was achieved by SHF, followed in order by NP and FFX, when the block size was small.
However, all protected models dropped the segmentation performance compared to the baseline.
We also confirmed that the protection model with FFX, which has a small block size, provided inaccurate results for unauthorized users who do not know the correct key or transformation.
As future work, we will improve the protected model's performance and evaluate robustness against various attacks.


\bibliographystyle{ieicetr}
\bibliography{strings}

\end{document}